\documentclass[prl,twocolumn,showpacs,superscriptaddress,amsmath,amssymb]{revtex4}

\usepackage{graphicx}
\usepackage{amssymb,amsmath}
\usepackage{dcolumn}
\usepackage{bm}
\usepackage{mathrsfs}
\usepackage{color}

\begin{document}

\title{Transmutation from Skyrmions to Half-Solitons\\
 Driven by the Nonlinear Optical Spin-Hall Effect}

\author{H. Flayac}
\affiliation{Institut Pascal, PHOTON-N2, Clermont Universit\'{e}, Blaise Pascal University, CNRS, 24 avenue des Landais, 63177 Aubi\`{e}re Cedex, France}
\author{D. D. Solnyshkov}
\affiliation{Institut Pascal, PHOTON-N2, Clermont Universit\'{e}, Blaise Pascal University, CNRS, 24 avenue des Landais, 63177 Aubi\`{e}re Cedex, France}
\author{I. A. Shelykh}
\affiliation{Science Institute, University of Iceland, Dunhagi-3, IS-107, Reykjavik,
Iceland}
\affiliation{Division of Physics and
Applied Physics, Nanyang Technological University 637371, Singapore}
\author{G. Malpuech}
\affiliation{Institut Pascal, PHOTON-N2, Clermont Universit\'{e}, Blaise Pascal University, CNRS, 24 avenue des Landais, 63177 Aubi\`{e}re Cedex, France}

\begin{abstract}
We show that the polarization domains generated in the linear optical spin-Hall effect by the analog of spin-orbit interaction for exciton-polaritons are associated with the formation of a Skyrmion lattice. In the non-linear regime, the spin anisotropy of the polariton-polariton interactions results in a spatial compression of the spin domains and in an abrupt transformation of the Skyrmions into half-solitons, associated with both the focalization of the spin currents and the emergence of a strongly anisotropic emission pattern.

\end{abstract}
\maketitle

\emph{Introduction}.
The ultimate goal of spintronics is to replace charge currents by pure spin currents. The spin-Hall effect is one of the key concepts in this research field, since it allows the formation of a spin current perpendicular to the electric current without a magnetic field. Originally proposed by Dyakonov and Perel' in 1971 \cite{Dyakonov} and rediscovered by Hirsch in 1999 \cite{Hirch}, it attracts a lot of attention now \cite{SHI2,SHI3}. Basically, this effect originates from the electron's spin-orbit interaction that makes its scattering on impurities spin-dependent. One of the most serious obstacles to the realization of spintronic devices is the dramatic role played by the processes of spin relaxation, inducing information losses. The optical counterpart of spintronics, namely spin-optronics, has been proposed as a valuable alternative, since the characteristic decoherence times of optical excitations are orders of magnitude longer than those of electrons and holes \cite{Spinoptronics}.

Exciton-polaritons (polaritons), being a mixture of quantum-well excitons and cavity photons, appear as the optimal candidates for the realization of optoelectronic devices showing a polarization-selective behavior. The large photonic fraction of polaritons allows an efficient resonant optical excitation with a laser pump at any energy, momentum and density, with a well defined spin to form macroscopically occupied and coherent states having a controllable spatial extension. Moreover, the polariton mass is five orders of magnitude smaller than that of free electron, allowing their propagation at large velocities over hundreds of microns \cite{Wertz} preserving the spin coherence \cite{NOSHE}. On the other side, their excitonic part allows them to interact efficiently with each others. These interacting photonic particles are therefore demonstrating an outstanding non-linear optical response \cite{OPO,Bistability}.

Polaritons have two spin projections on the structure growth axis of the sample \cite{ReviewSpin}. An analog of the spin-orbit interaction is present in microcavities, provided by the so-called TE-TM splitting \cite{Berry}. This feature gives birth to the optical spin-Hall effect (OSHE) \cite{OSHE} which became a popular research topic \cite{OSHENPhys,Langbein,OSHEOpt,AnisotropeOSHE}. The OSHE however remains a linear effect, which does not capitalize on the exceptional non-linear optical response of the polariton system \cite{OSHEOpt}. Furthermore, the created spin currents are by nature relatively weak, and demonstrate an enormous angular aperture of almost 45 degrees.

Polariton-polariton interactions are strongly spin anisotropic \cite{Renucci2005}, which imposes a spin-dependent nonlinear response \cite{ShelykhPRL,Multistability,MultistabilityExp}, resulting in the formation of spin polarized topological excitations namely half-vortices \cite{RuboHV,LagoudakisHV} and half-solitons \cite{FlayacHS,HivetHS}. These objects have circularly polarized dips on a linearly polarized background \cite{FlayacHV}, the density in the center being one half of that at infinity. They differ from the topological excitations in spin-isotropic condensates, which are Skyrmions or coreless vortices \cite{KasamatsuSkyrmion}, appearing as domains of circular polarization inversion, while the total density remains constant. It was recently understood \cite{SolnyshkovMonop} that half-integer topological defects in polariton condensates behave as stable magnetic charges accelerated by effective magnetic field. They can therefore be foreseen as ultrafast spinoptronics information mediators \cite{FlayacNJP}.

In this letter, we describe the spectacular transition arising between the linear and nonlinear regimes in the OSHE. In the linear regime, we demonstrate that the spin domains generated by the interplay between the particle flux and the spin-orbit like interaction are carrying topological charges, similar to Skyrmions forming a lattice. In the nonlinear regime one observes a dramatic change of the polarization pattern, emission profile and nature of the topological excitations. The spin domains are found to be spatially compressed by the spin-anisotropic polariton-polariton interaction. The phase transition occurs when the healing length of the nonlinear fluid becomes smaller than the size of the polarization domains. The Skyrmions created by the spin-orbit interaction abruptly transform into half-solitons, stabilized in turn by the interactions. Finally, the total emission profile switches from isotropic to strongly anisotropic, accompanied by a very strong focalization of the spin currents.

\begin{figure}[tbp]
\includegraphics[width=0.85\linewidth]{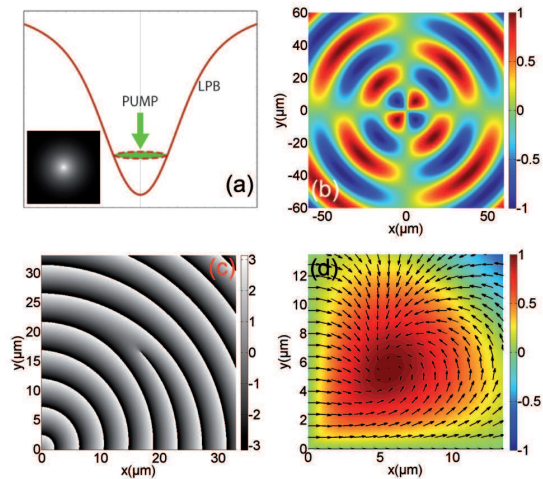}
\caption{(a) Pumping scheme showing the lower polariton branch (LPB) and its excitation on a ring by the pump spot. Inset: cylindrically symmetric distribution of the total intensity under the analytical description. (b) Degree of circular polarization $\rho_c$ showing the antisymmetric circular polarization domains. (c) Zoom on the phase of the $\sigma_+$ component characterized by a fork-like dislocation at the position of the density minimum which is the typical signature of a topological defect in this component. No dislocation occurs at the $\sigma_-$ minima (red regions of panel in (b)). (d) Pseudospin texture zooming on a Skyrmion and showing the winding of the in-plane pseudospin around its core (black arrows).}
\label{fig1}
\end{figure}

\emph{Linear OSHE}.
The original proposal for the OSHE \cite{OSHE} involved the resonant injection of exciton polaritons with a well-defined momentum and linear polarization within the disorder landscape of a microcavity. The resulting coherent Rayleigh scattering allowed a homogeneous redistribution of polaritons on an elastic ring in $k$-space. However, in modern microcavities the disorder tends to be strongly reduced and would anyway be screened by the particles interactions in the nonlinear regime. An alternative configuration for the observation of OSHE is to use a blue detuned laser with a small spot \cite{Langbein} giving a sufficient extension in momentum space to excite directly the polariton dispersion on a ring [see Fig.\ref{fig1}(a)]. The $\mathbf{k}$-dependent precession of the polariton pseudospin $\mathbf{S}$ around the effective magnetic field $\mathbf{H}_{LT}$ associated with the TE-TM splitting results in the appearance of alternating circularly polarized domains in the four quarters of the plane, as shown in Fig.\ref{fig1}(b).

For simplicity let us consider the ideal situation of particles continuously injected at $\mathbf{r}=\mathbf{0}$, linearly polarized along the $x$ direction and well defined energy. This boundary condition excites a specific linear combination of TE and TM eigenmodes for each polar angle. The resulting polariton wave function, stationary solution of the spin-dependent Schr\"{o}edinger equation with external pumping term written here on the $(x,y)$ polarization basis reads
\begin{eqnarray}
\label{WFTETM1}
{\psi _x}({{r}},\varphi ) &=& \sqrt{n_x}{e^{ - \frac{r}{{{r_0}}}}}\left( {{e^{i{k_{t}}r}}{{\cos }^2}\varphi  + {e^{i{k_{l}}r}}{{\sin }^2}\varphi } \right)\\
\label{WFTETM2}
\nonumber{\psi _y}({{r}},\varphi ) &=& \sqrt{n_y}{e^{ - \frac{r}{{{r_0}}}}}\left( {{e^{i{k_{t}}r}}\sin \varphi \cos \varphi  + {e^{i{k_{l}}r}}\sin \varphi \cos \varphi } \right)\\
\end{eqnarray}
Here ${r_0}=\hbar k\tau /{m^*}$ with $\tau$ and $m^*$ are the polariton lifetime and effective mass respectively and $k=(k_l+k_t)/2$. $k_{l,t}=\sqrt{2 m^* E_{l,t}}/\hbar$ where $E_{l,t}$ is the energy of the TM and TE mode respectively. The distribution of the total emission shown on the inset of the Fig.\ref{fig1}(a) is assumed here to be fully isotropic. Fig.\ref{fig1}(b) shows the degree of circular polarization $\rho_c=\Im \left( {\psi _x^*{\psi _y}} \right)/(n_x+n_y)$ of the polariton emission. It demonstrates the typical periodic polarization oscillations with maxima lying along diagonal directions, perfectly reproducing the polarization pattern of the original OSHE. The positions of the maxima are defined by
\begin{eqnarray}
\varphi  &=& \frac{\pi }{4},\frac{{5\pi }}{4}:{r_p^+ } = \frac{{\left( {2p - 1} \right)\pi }}{{2\left( {{k_t} - {k_l}} \right)}},{\rm{ }}{r_p^- } = \frac{{p\pi }}{{{k_t} - {k_l}}}\\
\varphi  &=& \frac{{3\pi }}{4},\frac{{7\pi }}{4}:{r_p^+ } = \frac{{p\pi }}{{{k_t} - {k_l}}},{\rm{ }}{r_p^- } = \frac{{\left( {2p - 1} \right)\pi }}{{2\left( {{k_t} - {k_l}} \right)}}
\end{eqnarray}
where $p\in\mathbb{Z}$ and the indices $\pm$ refer to the $\sigma_\pm$ component respectively given that $\psi_\pm=(\psi_x\pm i\psi_y)/\sqrt{2}$. A fundamental feature is that the cores of the spin domains are characterized by phase singularities in the circular polarized components shown in Fig.\ref{fig1}(c). They appear because the effective field projections on the $x$ axis are exactly opposite on the two sides of the diagonal direction $\varphi=p\pi/4$, inducing a $\pi$ phase difference after a half-period rotation of the pseudospin. This is the time-reversal of the half-vortex unwinding scheme \cite{Matthews}, which in our case produces not a half-vortex, but a Skyrmion excitation, characterized by a homogeneous total density distribution [see inset of Fig.\ref{fig1}(a)]. The latter appears because the \emph{spin-anisotropic} interactions, known to stabilize the half-vortices, are absent yet, and the system is governed instead by the \emph{spin-isotropic} effective magnetic energy. The pseudospin texture zoomed on a Skyrmion is shown in Fig.\ref{fig1}(d).

\emph{Nonlinear OSHE}.
When the density increases, the quantum fluid becomes non-linear, and can support topological excitations driven by the interparticle interactions, their size being defined by the healing length $\xi$. Due to the spin anisotropy of the polariton interactions, Skyrmions cannot appear any more and are replaced by half-solitons. Qualitatively, one can expect the transition to occur when the healing length becomes smaller than the Skyrmion size (of the order of $r_0^\pm$). Once transformed, the resulting defect starts to feel the effective magnetic force provided by the TE-TM splitting, which now acts as a weak perturbation.

The complete simulation of the problem requires numerical solutions of non-linear wave equations. We model the system of resonantly injected cavity polaritons with a system of equations for the coupled photonic $\phi(\mathbf{r},t)$ and excitonic fields $\chi(\mathbf{r},t)$, each of which have two components written here on the circular polarization basis:
\begin{eqnarray}
\label{PhiDyn}
\nonumber\frac{{\partial {\phi _ \pm }}}{{\partial t}} = &-&\frac{{{\hbar ^2}}}{{2{m_\phi }}}\Delta {\phi _ \pm } + {\Omega_R}{\chi _ \pm } + \beta {\left( {\frac{\partial }{{\partial x}} \mp i\frac{\partial }{{\partial y}}} \right)^2}{\phi _ \mp }\\
 &+& {P_ \pm }{e^{ - i{\omega _P}t}} - \frac{{i\hbar }}{{2{\tau _\phi }}}{\phi _ \pm } \\
\label{ChiDyn}
\nonumber\frac{{\partial {\chi _ \pm }}}{{\partial t}} = &-&\frac{{{\hbar ^2}}}{{2{m_\chi }}}\Delta {\chi _ \pm } + {\Omega_R}{\phi _ \pm }\\
 &+& \left( {{\alpha _1}{{\left| {{\chi _ \pm }} \right|}^2} + {\alpha _2}{{\left| {{\chi _ \mp }} \right|}^2}} \right){\chi _ \pm } - \frac{{i\hbar }}{{2{\tau _\chi }}}\chi
\end{eqnarray}
In the above expressions $V_R=5$ meV denotes half of the Rabi splitting coupling excitonic and photonic fields, $\tau_\chi=400$ ps and $\tau_\phi=10$ ps are the lifetimes of excitons and cavity photons respectively, $\mathbf{P}(\textbf{r},t)$ is a 2 $\mu$m large Gaussian, quasi-resonant and linearly polarized (along the $x$-axis) optical pump having a frequency $\omega_P$ blue detuned by $+1$ meV from the lower polariton branch. $m_\chi=0.4 m_0$, $m_\phi=5\times10^{-5}m_0$ are the effective masses of the excitons and cavity photons respectively ($m_0$ is the electron mass). The constant $\beta=\hbar^2/4(1/m_{\phi}^{TM}-1/m_{\phi}^{TE})$, where $m_{\phi}^{TE}=m_\phi$ and $m_{\phi}^{TM}=0.95m_\phi$ are the masses of the photonic TE and TM modes, is the strength of the photonic TE-TM splitting. The corresponding terms give rise to the in-plane effective magnetic field $\mathbf{H}_{LT}(\textbf{k})$. The constants $\alpha_1$ and $\alpha_2=-0.2\alpha_1$ describe the spin- anisotropic exciton- exciton interactions.

\begin{figure*}[tbp]
\includegraphics[width=0.85\linewidth]{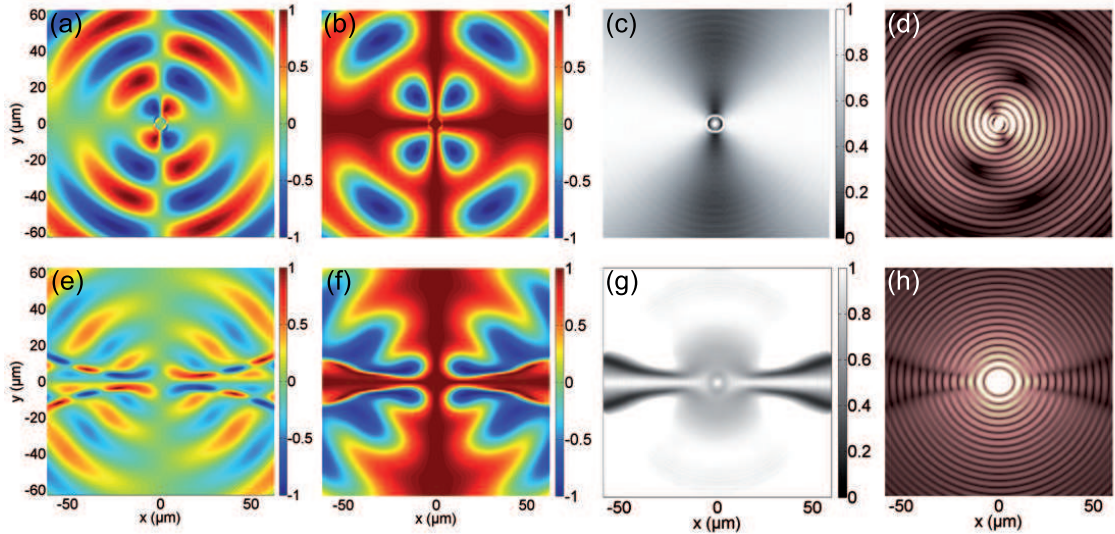}
\caption{(a-d) Linear and (e-h) nonlinear regimes. (a,e) Degree of circular polarization showing the compression of the domains and the half-soliton formation (e) while the density is increased. (b,f) Degree of linear polarization $\rho_l=(n_x-n_y)/(n_x+n_y)$. (b) the linear polarization oscillates in the oblique direction. (f) The half-solitons are clearly evidenced as domain walls between $x$ (red regions) and $y$ (blue regions) polarizations. (c,g) Total emission, normalized to compensate for the global decay, goes from almost homogeneous (c) towards the appearance of visible minima at the position of the half-solitons. (d,h) Interference patterns: In the linear regime (d) only the $\sigma_+$ component is plotted: typical forklike dislocations at the position of the density minima evidencing topological defects appearing in only one component. (c) The homogeneity of the total emission means that these defects are Skyrmions. (h) Phase of the total emission revealing an elongated phase slip along the total density minima evidencing further the half-solitons.}
\label{fig2}
\end{figure*}

The results are presented in Fig.\ref{fig2}: Panels (a) and (e) show the circular polarization degree in the linear and nonlinear regimes, correspondingly. In the linear regime one can notice that the spin pattern is not symmetric, with the spin domain being slightly compressed along the $y$-axis. Indeed, in a realistic simulation (and experiment) the use of a small size, monochromatic laser tends to excite TE and TM modes with two different wave vectors instead of two different energies as assumed in the original OSHE and to derive Eqs.(1,2). This induces two different velocities for the TE and TM propagating waves which would result in two different effective decay lengths. This is slightly shifting in (c) the directions of maximum circular polarization away from the diagonal direction and slightly breaks the homogeneous distribution of the total intensity. As soon as the polariton density increases so that the healing length $\xi\sim\hbar/\sqrt{\alpha_1 n_{\chi} m_\phi}$ becomes smaller than the size of the Skyrmions determined by $r_0^{\pm}$, the nonlinear effects start to dominate over the TE-TM splitting, and the interplay between the two leads to the transmutation (conversion) of Skyrmions into half-solitons. The latter are characterized by phase dislocations clearly visible close to the $x$-axis in (h). They also appear in (f) as a domain walls between $x$ and $y$ polarizations \cite{FlayacHS} and in (g) as sharp minima in the total density distribution. The angular dependence of the emission is therefore switching from smooth to strongly anisotropic. Interestingly one can see in (e) an oscillation of the circular polarization along the half soliton trajectory. Indeed, the decay of the total density far from the pump spot makes the half-solitons unstable against the effective field $\mathbf{H}_{LT}$ as soon as it exceeds the critical field $H_c=(\alpha_1-\alpha_2)n/4$ \cite{SolnyshkovMonop}. The two half-solitons of opposite circularity therefore get coupled, leading to spatial oscillations of the circular polarization degree. However, due to the finite life time of polaritons the integrated emission the solitons axes remains strongly circularly polarized, as visible in Fig.\ref{fig2}(b).

\begin{figure}[tbp]
\includegraphics[width=0.95\linewidth]{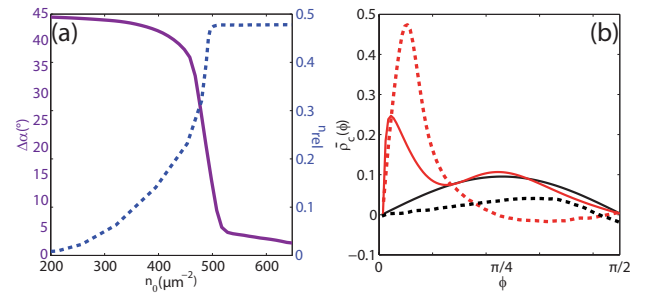}
\caption{(a) Phase transition: angular extension of the topological defects $\Delta\alpha$ (solid/purple line and left scale) and their relative contribution to the total density $n_{rel}$ (dashed/blue line and right scale). (b) Focalization of the spin currents: Analytical (solid lines) vs numerical results (dashed lines). Black (red) curves: spatially integrated degree of circular polarization $\overline \rho _c$ in the linear (nonlinear) regime.}
\label{fig3}
\end{figure}

In addition, we have extracted in Fig.\ref{fig3}(a) the angular extension $\Delta\alpha(\varphi)$ of the topological defects (solid/purple line) and their local contribution to the total density $n_{rel}=1-n_\phi^+/(n_\phi^++n_\phi^-)$ (dashed/blue line) as a function of the pump spot density $n_0$. The abrupt evolution at a critical density of both quantities demonstrate that we assist to a phase transition. Indeed the angular aperture of the topological defects switches from 40 degrees to less than 5 degrees and the relative density from 0 to almost 0.5 where it remains locked.

The \emph{compression} of the spin domains can be described in the framework of the pseudospin dynamics solely \cite{ReviewSpin}, which cannot address however the formation and dynamics of topological defects. The pseudospin precession equation reads $\partial_t {\mathbf S} = {\mathbf{H}} \times {\mathbf{S}}/\hbar$. The total effective magnetic field $\mathbf{H}$ represents the sum of the field provided by TE-TM splitting $\textbf{H}_{LT}$ lying in plane of the cavity and the effective field provided by the polariton-polariton interactions, responsible e.g. for the self-induced Larmor precession \cite{Solnyshkov2007} (intrinsic Zeeman splitting). The latter is directed perpendicular to the plane of the microcavity ($z$ direction) and describes the splitting between the circular polarized polariton states appearing due to the spin anisotropy of the interactions: $\textbf{H}_{{int}}=-(\alpha_1 - \alpha_2)\left(n_{+}-n_{-}\right)/2\textbf{u}_z$. The resulting system of nonlinear differential equations cannot be solved analytically in the general case.

If the pseudospin is initially aligned along the $x$-axis, important conclusions can be made studying the dynamics at short times. Indeed, if one considers the dynamics of the $S_y$ pseudospin projection (diagonal polarizations), it turns out that the \emph{contributions} of the TE-TM and nonlinear fields to this projection are able to cancel each other if the propagation direction angle $\varphi$ verifies the condition $H_0 \cos(2\phi)=-\alpha n$. This compensation determines the direction of the maximal oscillations of the polarization degree: instead of rotating around the in-plane field, the pseudospin turns further and further towards the $z$ axis, because its $S_y$ projection remains zero. In the linear case, this direction corresponds to the TE-TM field completely orthogonal to the initial pseudospin $\phi=p\pi/4$, while in the non-linear case the maximum is shifted towards the $y$ axis, along which the field is antiparallel to the injected pseudospin, until the size of the defects becomes smaller than $r_0^\pm$, when the phase transition occurs.

The focalization of the spin currents is illustrated in Fig.\ref{fig3}(b) showing the analytical (solid lines) and numerical (dashed lines) space-integrated degree of circular polarization ${\overline \rho _c}\left(\varphi\right) = \int {dr({n_ + } - {n_ - })/({n_ + } + {n_ - })}$. A substantial compression of the spin domains is obtained. In the linear regime (black lines) we see that the spin currents are almost unusable while in the nonlinear regime (red lines) the focusing is promising for a real spin current generator. The focalization however is much smaller in the analytical case than in the full Gross-Pitaevskii model [Eqs.(\ref{PhiDyn},\ref{ChiDyn})], shown in Fig.\ref{fig3}(b), and which takes into account the formation and dynamics of topological defects.

\emph{Conclusions}.
In summary, we have found that the linear optical spin-Hall effect is associated with the formation of a Skyrmion lattice. In the nonlinear regime a transmutation towards pairs of half-solitons occurs. The latter are associated with the breaking of the cylindrical symmetry of the total density and the strong focusing the spin currents in the real space.

\emph{Acknowledgments}.
We would like to thank M. Glazov and T. C. H. Liew for fruitful discussions. We acknowledge the support of the joint CNRS-RFBR PICS project, FP7 ITN Spin-Optronics (237252) and IRSES "POLAPHEN" project. I. A. S. benefited from Rannis "Center of Excellence in polaritonics" and IRSES "SPINMET" project.

\end{document}